\documentclass{article}
\usepackage[preprint]{spconf}
\usepackage{spconf,amsmath,amssymb,graphicx,booktabs,multirow}
\usepackage[T1]{fontenc}
\usepackage{microtype}
\usepackage{needspace,balance}
\usepackage{tikz}
\usepackage{float}
\copyrightnotice{\copyright\ IEEE 2026}

\usetikzlibrary{arrows.meta,positioning,calc,plotmarks,decorations.pathreplacing}
\usepackage[hidelinks]{hyperref}
\hypersetup{pdftitle={AnchorPrompt: Self-Distilled Soft Prompts for Robust Audio-Language Models},pdfauthor={},pdfsubject={ICASSP 2027 coauthor draft},pdfkeywords={audio-language models, soft prompts, adversarial robustness, hallucination}}
\newcommand{\method}{AnchorPrompt}

\newcommand{\ind}{\mathbf{1}}

\title{AnchorPrompt: \\Self-Distilled Soft Prompts for Robust Audio-Language Models}
\name{Pooneh Mousavi$^{1,2}$ \quad
  Amir Ivry$^{3}$ \quad Mirco Ravanelli$^{1,2}$ \quad
  Cem Subakan$^{1,2,4}$%
}
\address{
  $^{1}$Concordia University  \quad
  $^{2}$Mila -- Quebec AI Institute \quad
$^{3}$Technion -- IIT \quad
  $^{4}$Laval University}
\begin{document}
\ninept
\maketitle

\begin{abstract}
Large audio-language models (LALMs) are sensitive to input perturbations, such as noise, waveform corruption, and adversarial injections. We propose \method, an efficient adaptation method that keeps the model frozen and learns a single block of prompt vectors inserted at the decoder input, between the audio and question embeddings. We train these vectors through self-distillation over diverse audio and text perturbations. To improve answer consistency and mitigate hallucination, we use the model's prediction on the clean recording as the target for answerable inputs, and assign a refusal target when the audio lacks sufficient evidence to answer. Furthermore, \method\ is perturbation-agnostic at inference, requiring no prior detection of perturbations and enabling zero-shot transfer to unseen distortions. We evaluate three LALMs across three benchmarks and show that \method\ improves answer consistency in most tested conditions. Clean accuracy improves in six of nine model–benchmark pairs, with minimal impact on the remainder of $1.2$\% at most. Crucially, \method\ reduces hallucinations under severe audio corruption while keeping false refusals on clean audio rare. Finally, these consistency gains transfer to unseen perturbations, such as choice permutations and reverberation.\footnote{Audio examples and code: \url{https://poonehmousavi.github.io/anchorprompt}}
\end{abstract}

\section{Introduction}
\label{sec:intro}
\suppressfloats[t]
\begin{figure}[t]
\centering
\resizebox{\linewidth}{!}{
\begin{tikzpicture}[x=1cm,y=1cm,>=Latex,font=\fontsize{9}{10.5}\selectfont,
 box/.style={draw=black!60,rounded corners=2pt,align=center,inner sep=4pt,line width=.5pt},
 frozen/.style={box,fill=black!4},edge/.style={->,line width=.6pt},
 group/.style={draw=black!45,dashed,rounded corners=4pt,line width=.6pt},
 grouplabel/.style={fill=white,inner sep=1.5pt,font=\fontsize{8.5}{9.5}\selectfont\bfseries}]
\newcommand{\flameicon}[1]{\begin{scope}[shift={#1},scale=0.16]
 \fill[orange!90!red] (0,0) .. controls (-1.1,0.4) and (-1.0,1.6) .. (-0.2,2.2) .. controls (-0.1,1.5) and (0.3,1.3) .. (0.4,1.0) .. controls (0.9,1.5) and (1.1,0.5) .. (0,0) -- cycle;
 \fill[yellow!85!orange] (0,0.1) .. controls (-0.5,0.4) and (-0.4,1.0) .. (0,1.3) .. controls (0.45,1.0) and (0.5,0.4) .. (0,0.1) -- cycle;
\end{scope}}
\newcommand{\snowicon}[1]{\begin{scope}[shift={#1},scale=0.1,line width=0.55pt,cyan!70!blue]
 \foreach \a in {0,60,120} {\draw[rotate=\a] (-1,0) -- (1,0); \draw[rotate=\a] (0.55,0) -- (0.85,0.3) (0.55,0) -- (0.85,-0.3) (-0.55,0) -- (-0.85,0.3) (-0.55,0) -- (-0.85,-0.3);}
\end{scope}}
\useasboundingbox (-.45,-.4) rectangle (11.1,5.27);
\node[box,minimum width=2.15cm] (clean) at (1.1,4.4) {Clean audio\\+ question};
\node[frozen,minimum width=2.3cm,minimum height=0.8cm] (teacher) at (5.0,4.4) {\textbf{Frozen LALM}};
\node[align=center] (target) at (9.4,4.4) {Clean answer\\target $y_t$};
\draw[edge] (clean)--(teacher);\draw[edge] (teacher)--(target);
\draw[group] (-0.4,3.84) rectangle (10.6,5.04);
\node[grouplabel,anchor=west] at (-0.25,5.04) {Teacher};
\node[box,minimum width=1.5cm] (qin) at (0.72,2.5) {Perturbed\\question};
\node[box,minimum width=1.5cm] (ain) at (0.72,1.05) {Perturbed\\audio};
\node[frozen,minimum width=1.55cm] (temb) at (2.62,2.5) {Token\\embedding};
\node[frozen,minimum width=1.55cm] (aenc) at (2.62,1.05) {Audio\\encoder};
\draw[edge] (qin)--(temb);\draw[edge] (ain)--(aenc);
\def\sq{0.3}\def\sy{1.7}
\tikzset{lab/.style={font=\fontsize{7.5}{8.5}\selectfont,black!70}}
\foreach \x in {4.1,4.48,4.86} \draw[fill=black!18,draw=black!55] (\x-\sq/2,\sy-\sq/2) rectangle (\x+\sq/2,\sy+\sq/2);
\foreach \x in {5.4,5.78,6.54} \draw[fill=blue!35,draw=blue!60!black] (\x-\sq/2,\sy-\sq/2) rectangle (\x+\sq/2,\sy+\sq/2);
\node at (6.16,\sy) {$\cdots$};
\foreach \x in {7.2,7.58,7.96} \draw[fill=green!25,draw=green!45!black] (\x-\sq/2,\sy-\sq/2) rectangle (\x+\sq/2,\sy+\sq/2);
\draw[group] (3.8,1.38) rectangle (8.26,2.02);
\node[lab] at (4.48,2.17) {audio};
\draw[decorate,decoration={brace,amplitude=3pt},blue!60!black] (5.25,1.9)--(6.69,1.9);
\node[lab,text=blue!50!black] at (5.97,2.19) {$P$ ($L$ vectors)};
\node[lab] at (7.58,1.24) {question};
\node[lab,font=\fontsize{7.5}{8.5}\selectfont\itshape,fill=white,inner sep=1pt,anchor=west] at (3.9,1.38) {embedding space};
\draw[edge] (temb.east) -| (7.58,2.02);
\draw[edge] (aenc.east) -| (3.55,\sy) -- (3.8,\sy);
\node[frozen,minimum width=1.8cm,minimum height=0.8cm] (dec) at (9.4,\sy) {\textbf{Frozen LLM}\\\textbf{decoder}};
\draw[edge] (8.26,\sy)--(dec);
\draw[group] (-0.4,0.5) rectangle (10.6,3.3);
\node[grouplabel,anchor=west] at (-0.25,3.3) {Student};
\node[box,align=center] (loss) at (9.0,-0.02) {Self-distillation loss};
\draw[edge] (dec)--(dec|-loss.north);
\draw[edge,dashed,rounded corners=3pt] (target.east) -- (11.0,4.4) -- (11.0,-0.02) -- (loss.east);
\draw[edge,blue!60!black] (loss.west) -| node[pos=.85,left,black,font=\fontsize{8}{9}\selectfont] {Update $P$ only} (6.54,1.38);
\snowicon{(teacher.north east)} \snowicon{(dec.north east)} \snowicon{(temb.north east)} \snowicon{(aenc.north east)}
\flameicon{(6.88,2.07)}
\end{tikzpicture}
}
\caption{\textbf{Overview of \method{} framework.} A student LALM on perturbed input is distilled from a frozen teacher on clean input. When audio lacks sufficient clues to answer, the target is set to \texttt{CANNOT DETERMINE} to prevent hallucination. A soft prompt $P$ ($L$ learned vectors), inserted between audio and question embeddings, is the only trainable parameter. At inference, the student runs independently with $P$.}
\label{fig:method}
\end{figure}
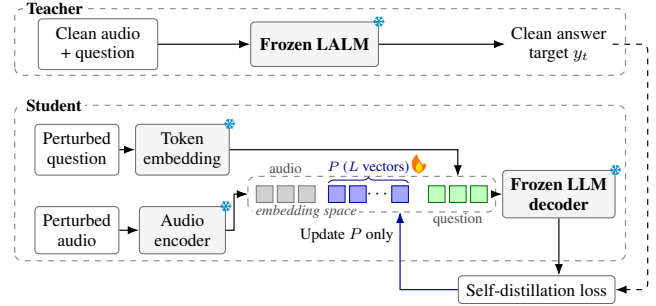

Large audio-language models (LALMs) combine an audio encoder with a large language model (LLM) to perform understanding and reasoning across speech, environmental sound, and music~\cite{tang2024salmonn,chu2024qwen2audio,xu2025qwen25,kimiteam2025kimiaudio,goel2025af3,xu2025qwen3,Bonzi2026_reasoning}. Benchmarks such as AIR-Bench~\cite{yang2024airbench} and Dynamic-SUPERB~\cite{huang2025dynamicsuperb} evaluate general audio tasks. Other benchmarks like MMAU~\cite{sakshi2025mmau}, MMAR~\cite{ma2025mmar}, and SAKURA~\cite{yang2025sakura} test expert knowledge, multi-step reasoning, and audio-grounded logic. However, these benchmarks primarily rely on clean audio recordings and static prompt templates. Consequently, they do not systematically evaluate model behavior under real-world conditions, where inputs are frequently noisy, incomplete, or subject to adversarial manipulation.

LALM robustness is essential in safety-critical domains like healthcare and emergency services~\cite{li2026audiotrust}. In practice, inputs face two main types of perturbations. First, \emph{natural perturbations} like background noise, missing segments, or reordered choices occur without malicious intent. Second, \emph{adversarial perturbations} like injected text or audio are designed to mislead the model. Recent studies show that current LALMs are vulnerable to both types. Performance drops significantly when noise or masking increases, even when the answer remains audible~\cite{mousavi2026faithfulness}. LALMs are also sensitive to choice permutations~\cite{lopez2026robustness} and audio compression~\cite{ivry2026task}. Furthermore, models often prioritize misleading text over audio evidence~\cite{wang2025disagree} and are easily derailed by injected voices~\cite{mousavi2026faithfulness, hou2025injection}. When given evidence-free audio, such as pure silence or heavy noise, LALMs tend to hallucinate instead of expressing uncertainty~\cite{kuan2025hear,zhao2026halluaudio,mousavi2025listen,kalai2025hallucinate,Wang2025CalmWhisper}. Recent evaluation frameworks highlight similar weaknesses in overall model reliability and safety~\cite{li2026audiotrust,ivry2026lalm}.

Text-only LLMs show similar sensitivity to typos, paraphrases, and choice ordering~\cite{zhu2023promptrobust,zheng2024mcq}. These issues can be mitigated via prompt optimization~\cite{zhou2023ape,pryzant2023protegi,wang2024promptagent,shinn2023reflexion,zhao2024expel,mu2025rop}. For LALMs, however, robustness also requires handling distortions in the audio modality. Noise, masking, or injected speech alter the audio embeddings supplied to the decoder. Standard hand-written instructions operate purely in text, so they cannot re-align these corrupted audio representations. In practice, hand-written instructions alone fail to stabilize model behavior across diverse audio and text perturbations.

To address this, we propose \method, an efficient soft-prompting framework. It inserts a small set of trainable soft-prompt vectors~\cite{lester2021power} between audio and question embeddings at the input of a frozen LALM decoder. We train these vectors via self-distillation in a teacher-student setup. The teacher is the frozen LALM operating on clean audio. The student is the same model processing perturbed inputs with the soft prompt. The student learns to replicate the teacher's clean prediction. When the audio lacks sufficient evidence to answer, the student learns to output a refusal target, preferring to refuse rather than hallucinate. Unlike prior methods that fine-tune model parameters~\cite{lin2026echodistill}, train task-specific adapters~\cite{kuan2025listen}, or modify decoding mechanics~\cite{hsu2025aad,lin2026silence}, \method\ requires zero model parameter updates and relies on self-distillation instead of ground-truth labels. Furthermore, \method\ is perturbation-agnostic at inference. It requires no prior detection of distortions and enables zero-shot transfer to unseen perturbations and attacks.

We evaluate \method\ across three LALMs and three benchmarks covering diverse natural and adversarial perturbations. We compare our approach against hand-written instruction baselines, assessing clean accuracy, distortion robustness, and hallucination mitigation. We also analyze zero-shot transfer to perturbations not seen during training, such as choice permutations and room reverberation.
\section{Method}
\label{sec:method}
\begin{figure}[!t]
\centering
\resizebox{\linewidth}{!}{
\begin{tikzpicture}[x=0.5cm,y=0.09cm,font=\fontsize{8}{9}\selectfont]
\definecolor{rp0}{HTML}{126B92}
\definecolor{rp1}{HTML}{C86520}
\definecolor{rp2}{HTML}{5C4B88}
\foreach \v/\t in {0/70,10/80,20/90,30/100}{
  \draw[gray!20] (0.0,\v) -- (6.5,\v);
  \node[anchor=east,inner sep=1pt] at (-0.05,\v) {\t};
}
\draw[gray!70] (0.0,30) -- (0.0,0) -- (6.5,0);
\node at (0.6,-4) {1};
\node at (2.4,-4) {4};
\node at (4.2,-4) {8};
\node at (6.0,-4) {16};
\node at (3.25,-9) {\textbf{(a)} prompt length $L$ (vectors)};
\draw[rp0,line width=0.8pt] plot[mark=*,mark size=1.4pt] coordinates {(0.600,11.38) (2.400,12.92) (4.200,13.25) (6.000,13.67)};
\draw[rp0,line width=0.6pt,dashed] plot[mark=*,mark size=1.1pt,mark options={fill=white}] coordinates {(0.600,27.03) (2.400,27.42) (4.200,28.36) (6.000,28.52)};
\draw[rp1,line width=0.8pt] plot[mark=square*,mark size=1.4pt] coordinates {(0.600,7.23) (2.400,10.13) (4.200,12.39) (6.000,10.38)};
\draw[rp1,line width=0.6pt,dashed] plot[mark=square*,mark size=1.1pt,mark options={fill=white}] coordinates {(0.600,16.80) (2.400,22.13) (4.200,26.22) (6.000,22.77)};
\draw[rp2,line width=0.8pt] plot[mark=triangle*,mark size=1.4pt] coordinates {(0.600,3.83) (2.400,4.84) (4.200,6.86) (6.000,7.74)};
\draw[rp2,line width=0.6pt,dashed] plot[mark=triangle*,mark size=1.1pt,mark options={fill=white}] coordinates {(0.600,23.71) (2.400,24.03) (4.200,24.66) (6.000,27.17)};
\foreach \v in {0,10,20,30}{ \draw[gray!20] (7.2,\v) -- (13.7,\v); }
\draw[gray!70] (7.2,30) -- (7.2,0) -- (13.7,0);
\node at (7.80,-4) {10};
\node at (9.15,-4) {20};
\node at (10.50,-4) {30};
\node at (11.85,-4) {50};
\node at (13.20,-4) {full};
\node at (10.45,-9) {\textbf{(b)} training items per dataset $K$};
\draw[rp0,line width=0.8pt] plot[mark=*,mark size=1.4pt] coordinates {(7.800,3.15) (9.150,4.79) (10.500,10.03) (11.850,10.16) (13.200,13.25)};
\draw[rp0,line width=0.6pt,dashed] plot[mark=*,mark size=1.1pt,mark options={fill=white}] coordinates {(7.800,21.88) (9.150,22.11) (10.500,28.91) (11.850,25.70) (13.200,28.36)};
\draw[rp1,line width=0.8pt] plot[mark=square*,mark size=1.4pt] coordinates {(7.800,7.36) (9.150,7.86) (10.500,10.25) (11.850,9.62) (13.200,12.39)};
\draw[rp1,line width=0.6pt,dashed] plot[mark=square*,mark size=1.1pt,mark options={fill=white}] coordinates {(7.800,9.56) (9.150,13.65) (10.500,20.88) (11.850,22.77) (13.200,26.22)};
\draw[rp2,line width=0.8pt] plot[mark=triangle*,mark size=1.4pt] coordinates {(7.800,2.08) (9.150,1.19) (10.500,7.11) (11.850,4.84) (13.200,6.86)};
\draw[rp2,line width=0.6pt,dashed] plot[mark=triangle*,mark size=1.1pt,mark options={fill=white}] coordinates {(7.800,18.05) (9.150,17.10) (10.500,22.77) (11.850,25.91) (13.200,24.66)};
\node[rotate=90] at (-1.05,15) {consistency / refusal (\%)};
\draw[rp0] plot[mark=*,mark size=1.4pt] coordinates {(0.3,35)};
\node[anchor=west,inner sep=1pt] at (0.45,35) {SAKURA};
\draw[rp1] plot[mark=square*,mark size=1.4pt] coordinates {(2.9,35)};
\node[anchor=west,inner sep=1pt] at (3.05,35) {MMAU};
\draw[rp2] plot[mark=triangle*,mark size=1.4pt] coordinates {(5.2,35)};
\node[anchor=west,inner sep=1pt] at (5.35,35) {MMAR};
\draw[black!70,line width=0.8pt] (7.6,35) -- (8.2,35); \node[anchor=west,inner sep=1pt] at (8.3,35) {consistency};
\draw[black!70,line width=0.6pt,dashed] (11.0,35) -- (11.6,35); \node[anchor=west,inner sep=1pt] at (11.7,35) {refusal};
\end{tikzpicture}
}
\caption{Qwen2.5-Omni validation sweeps: clean-answer consistency on perturbed items assigned clean-answer targets (solid) and refusal under the two refusal-target conditions (dashed) against (a) prompt length with full data and (b) training items per dataset at $L=8$. One seed per point.}
\vspace{-.5cm}
\label{fig:scaling}
\end{figure}
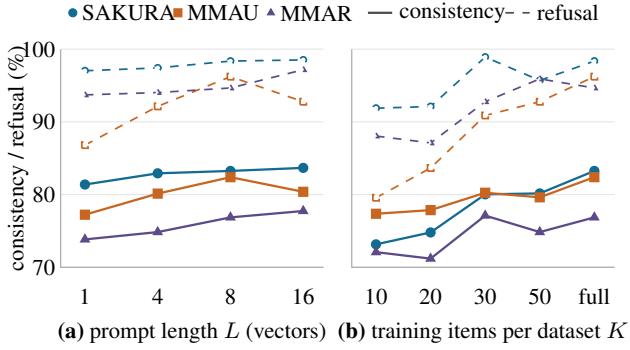

Fig.~\ref{fig:method} gives an overview of \method. We add a soft prompt in the form of a small set of learnable vectors to the decoder input of a frozen LALM. The rest of the model parameters remain frozen, and we only train the newly added vectors. Training follows a self-distillation setup, in which a teacher and a student share the same frozen LALM. The teacher sees a clean input and provides the target answer. The student sees a perturbed version of the same input with the soft prompt and learns to stay consistent with the teacher's answer, or to decline under the two refusal-target conditions defined in Sec.~\ref{sec:targets}. At inference, we use only the student. Sec.~\ref{sec:prompt} describes the soft prompt, Sec.~\ref{sec:targets} the training inputs and their targets, and Sec.~\ref{sec:objective} the training objective.

\subsection{Soft prompt}
\label{sec:prompt}
An LALM receives a recording, a question, and its answer options. The audio encoder turns the recording into a sequence of audio embeddings, and the language model decoder reads these embeddings jointly with the embedded question and options. We keep all model weights frozen and add a soft prompt $P\in\mathbb{R}^{L\times D}$, which consists of $L=8$ learnable vectors with the same dimension $D$ as the token embeddings of the language model.
We insert the soft prompt into the decoder’s input embedding sequence, between the audio embeddings and the embedded question, and leave the rest of the input unchanged.
We initialize each vector with the embedding of a randomly chosen vocabulary token. For a given LALM, a single soft prompt is shared across all audio perturbations. When evaluating multiple LALMs, each model is trained and evaluated with its own dedicated soft prompt.

\subsection{Training inputs and targets}
\label{sec:targets}
\begin{table*}[!t]
\centering
\caption{Test results (\%): accuracy~/~clean-answer consistency (defined in Sec.~\ref{sec:eval}). A refusal counts as incorrect; a question is consistent when the setting keeps its clean-input behavior, i.e.\ selects the same option or refuses on both inputs. Noise pools 10 and 0~dB SNR; Mask pools 40 and 60\% masking; Audio inj.\ is adversarial audio injection (SAKURA only). Manual is the hand-written instruction at the soft-prompt position.}
\label{tab:main}
\resizebox{0.8\textwidth}{!}{
\renewcommand{\arraystretch}{0.93}\begin{tabular*}{\textwidth}{@{\extracolsep{\fill}}lllrrrrr@{}}
\toprule
 & & & Clean & \multicolumn{4}{c}{Accuracy / clean-answer consistency} \\
\cmidrule(l){5-8}
Model & Benchmark & Prompt & acc. & Noise & Mask & Text inj. & Audio inj. \\
\midrule
Qwen2.5 & SAKURA & Base & 65.9 & 58.8/81.0 & 58.8/78.2 & 27.4/37.7 & 29.5/38.7 \\
 &  & Manual & 60.6 & 52.1/81.2 & 52.1/79.0 & 17.2/29.1 & 26.8/37.7 \\
 &  & \textbf{AnchorPrompt} & \textbf{75.1} & \textbf{68.6}/\textbf{84.5} & \textbf{68.5}/\textbf{83.1} & \textbf{73.2}/\textbf{87.6} & \textbf{68.2}/\textbf{76.3} \\
 & MMAU & Base & 73.4 & 67.8/84.2 & 65.6/80.4 & 31.1/43.1 & -- \\
 &  & Manual & 68.1 & 60.3/84.3 & 56.6/79.8 & 21.2/31.2 & -- \\
 &  & \textbf{AnchorPrompt} & \textbf{73.3} & \textbf{67.2}/\textbf{84.3} & \textbf{65.6}/\textbf{82.0} & \textbf{66.9}/\textbf{81.7} & -- \\
 & MMAR & Base & 54.2 & 46.2/76.3 & 39.9/64.8 & 16.8/36.9 & -- \\
 &  & Manual & 43.3 & 35.6/77.5 & 28.7/66.4 & 10.2/31.4 & -- \\
 &  & \textbf{AnchorPrompt} & \textbf{62.3} & \textbf{56.6}/\textbf{78.6} & \textbf{52.7}/\textbf{72.0} & \textbf{54.0}/\textbf{75.0} & -- \\
\midrule
Qwen3 & SAKURA & Base & 73.0 & 68.9/84.4 & 63.3/76.6 & 36.8/47.3 & 40.0/47.9 \\
 &  & Manual & 69.2 & 65.8/84.4 & 59.9/77.1 & 35.0/46.8 & 38.4/47.0 \\
 &  & \textbf{AnchorPrompt} & \textbf{79.3} & \textbf{78.1}/\textbf{88.6} & \textbf{74.0}/\textbf{83.2} & \textbf{80.2}/\textbf{90.3} & \textbf{73.6}/\textbf{79.3} \\
 & MMAU & Base & 74.8 & 67.2/81.6 & 63.8/76.3 & 40.6/51.2 & -- \\
 &  & Manual & 74.1 & 65.5/80.4 & 61.7/75.6 & 40.3/50.6 & -- \\
 &  & \textbf{AnchorPrompt} & \textbf{76.7} & \textbf{74.7}/\textbf{86.7} & \textbf{72.3}/\textbf{82.6} & \textbf{76.9}/\textbf{88.8} & -- \\
 & MMAR & Base & 62.0 & 50.2/72.4 & 44.8/63.3 & 29.2/42.1 & -- \\
 &  & Manual & 61.5 & 49.8/74.8 & 43.2/62.5 & 32.2/47.1 & -- \\
 &  & \textbf{AnchorPrompt} & \textbf{69.5} & \textbf{66.1}/\textbf{85.1} & \textbf{59.9}/\textbf{75.8} & \textbf{66.1}/\textbf{81.6} & -- \\
\midrule
AF3 & SAKURA & Base & 68.2 & 61.4/77.8 & 60.0/81.2 & 8.2/27.0 & 21.8/40.0 \\
 &  & Manual & 68.0 & 63.3/78.5 & 60.5/81.6 & 7.3/26.0 & 22.5/38.7 \\
 &  & \textbf{AnchorPrompt} & \textbf{67.0} & \textbf{64.3}/\textbf{81.5} & \textbf{60.8}/\textbf{84.8} & \textbf{67.2}/\textbf{78.5} & \textbf{52.1}/\textbf{68.3} \\
 & MMAU & Base & 75.5 & 73.6/87.4 & 72.7/88.0 & 29.7/40.5 & -- \\
 &  & Manual & 77.0 & 73.2/87.1 & 73.4/88.0 & 30.0/39.4 & -- \\
 &  & \textbf{AnchorPrompt} & \textbf{74.4} & \textbf{71.4}/\textbf{86.6} & \textbf{72.3}/\textbf{89.1} & \textbf{72.2}/\textbf{87.8} & -- \\
 & MMAR & Base & 57.8 & 53.8/82.6 & 50.6/78.9 & 13.0/30.5 & -- \\
 &  & Manual & 57.0 & 53.9/83.1 & 51.0/80.3 & 10.7/28.4 & -- \\
 &  & \textbf{AnchorPrompt} & \textbf{59.2} & \textbf{53.9}/\textbf{83.0} & \textbf{53.9}/\textbf{81.4} & \textbf{49.6}/\textbf{69.5} & -- \\
\bottomrule
\end{tabular*}
}
\end{table*}
For each training item $x$, the teacher answers the clean input once with greedy decoding. Our goal is consistency: under a perturbation, the model should give the same answer as on the clean input. We therefore use the teacher's answer as the target, even when it differs from the ground-truth answer. We then create perturbed versions $\tilde x$ of the item with the perturbations of Sec.~\ref{sec:perturb} and assign each input a target $y_t$.
The inputs fall into three groups:
\begin{itemize}\setlength{\itemsep}{0pt}
\item \textbf{Clean input}, where $y_t$ is the teacher's answer. Training on clean inputs encourages the soft prompt to preserve the teacher's clean-input answers.
\item \textbf{Perturbations with clean-answer targets}, where $y_t$ is the teacher's answer. This group contains noise and masking below the extreme levels, adversarial audio injection, and adversarial text injection. The student is trained to stay consistent with the teacher under these perturbations.
\item \textbf{Refusal-target conditions}, where the target \(y_t\) is the phrase CANNOT DETERMINE. This group contains 100\% masking and noise at $-20$ dB signal-to-noise ratio (SNR). The student is trained to decline under these two conditions.
\end{itemize}

\subsection{Training objective}
\label{sec:objective}
Let $\mathcal S$ be the set of all training pairs $(\tilde x, y_t)$, where $\tilde x$ is a clean or perturbed input and $y_t$ is the clean-answer or refusal target assigned in Sec.~\ref{sec:targets}. The student is the frozen LALM with the soft prompt $P$. As in standard soft-prompt tuning~\cite{lester2021power}, we minimize the cross-entropy of the student on the target tokens, so that the student learns the assigned target. 
Only the soft prompt is updated, and the teacher's answers are computed once before training. Ground-truth answers are excluded from the loss calculation and used only for evaluation and for constructing adversarial injections using the dataset from~\cite{mousavi2026faithfulness}.

\section{Experimental Setup}
\label{sec:setup}
\subsection{Data and partitioning}
We use SAKURA~\cite{yang2025sakura}, MMAU test-mini~\cite{sakshi2025mmau}, and MMAR~\cite{ma2025mmar}. SAKURA contains 4{,}000 multiple-choice questions over 2{,}000 recordings in four tracks (animal, emotion, gender, and language), and MMAU and MMAR contain about 1{,}000 questions each over speech, sound, and music. We split the recordings of each track into 20\% training, 16\% validation, and 64\% test. The test sets contain 2{,}560 SAKURA, 640 MMAU, and 637 MMAR questions, and the training set contains 1{,}200 questions. These counts are before applying perturbations.
\vspace{-.3cm}
\subsection{Perturbations}
\label{sec:perturb}
We adopt perturbations such as noise, masking, and adversarial audio injections from Mousavi et al.~\cite{mousavi2026faithfulness}.
Noise is white Gaussian noise at 20, 10, 0, $-10$, or $-20$~dB SNR, and masking sets randomly chosen 100~ms chunks of the recording to zero until 20, 40, 60, 80, or 100\% of the recording is silent. 
We designate $-20$~dB SNR noise and 100\% masking as refusal-target conditions. While full masking completely eliminates the original audio evidence, $-20$ dB SNR noise heavily degrades it, making a correct answer highly unlikely, though not necessarily impossible, depending on the question. We report the 10 and 0 dB SNR and 40\% and 60\% masking results in Table~\ref{tab:main}, and the two refusal-target conditions in Table~\ref{tab:halluc}.
For adversarial audio injection, we reuse the recordings of~\cite{mousavi2026faithfulness}, in which a synthesized voice speaks a wrong option inside the recording at a power at or below that of the source. We apply it only on the SAKURA dataset, whose tasks are mostly identification, so an injected word does not change the content. On MMAU and MMAR, which require reasoning over the content, an injected word could change the context and therefore the answer. For adversarial text injection, we write seven templates that add a sentence asserting a randomly chosen wrong option, similar to the conflicting text cues studied by Wang et al.~\cite{wang2025disagree}. An example is \emph{``Context: this recording is believed to contain crow.''}, which is prepended to the question. For zero-shot transfer (Sec.~\ref{sec:ood}), we also evaluate two out-of-distribution (OOD) perturbations: a permutation of the answer choices and reverberated audio with a reverberation time of 1~s.

\subsection{Models, baselines, and training}
\label{sec:models}
We use Qwen2.5-Omni-7B (Qwen2.5)~\cite{xu2025qwen25}, Qwen3-Omni-30B-A3B (Qwen3)~\cite{xu2025qwen3}, and Audio Flamingo~3 (AF3)~\cite{goel2025af3}. We evaluate each LALM with a fixed multiple-choice prompt, and our clean accuracy on MMAU and MMAR is close to the accuracy reported by the model authors.

\textbf{Baselines.} \texttt{Base} is the frozen LALM with its original prompt. \texttt{Manual} adds the following hand-written instruction at the position of the soft prompt: \emph{``Base your answer on the audible evidence in the recording. Answer when that evidence supports an option, even if the recording is noisy or incomplete. If the recording provides insufficient evidence to distinguish the options, respond with CANNOT DETERMINE. Otherwise, output only the selected option's letter.''}

\textbf{Training.} We choose the prompt length and the training-set size on the Qwen2.5-Omni validation split (Fig.~\ref{fig:scaling}). Consistency and refusal level off at $L=8$ and keep improving with more training data, so we use $L=8$ and the full training set for all models. We train the soft prompt for three epochs with AdamW and a learning rate of $10^{-3}$, and we select the checkpoint with the highest agreement with the teacher on held-out validation recordings. The soft prompt has 28{,}672 trainable parameters for Qwen2.5-Omni and AF3 and 16{,}384 for Qwen3-Omni, which is less than 0.001\% of the parameters of each model. Training takes 3.3~h for Qwen2.5-Omni and 2.7~h for AF3 on a single NVIDIA L40S (48~GB), and 8.8~h for Qwen3-Omni on a single NVIDIA A100 (80~GB).

\subsection{Evaluation}
\label{sec:eval}
We compare three settings: Base, Manual, and \method. All models decode greedily. We report the following metrics.

\textbf{Accuracy} is the percentage of questions answered with the ground-truth option. A refusal counts as incorrect.

\textbf{Clean-answer consistency} measures whether a model maintains its original selection when the input is perturbed. For each test question, we record the option $y_{\mathrm{clean},i}$ selected for the clean question $x_i$ and the option $y_{\mathrm{pert},i}$ selected for the perturbed input $\tilde{x}_i$. Consistency is the percentage of questions for which the two selections match, treating a refusal as an additional option:  $\mathrm{Consistency}=\frac{100}{N}\sum_{i=1}^{N}\ind\big[\,y_{\mathrm{pert},i}=y_{\mathrm{clean},i}\,\big]$, where $N$ is the number of test questions.
We always compute consistency against the clean answer of the specific configuration being evaluated, rather than the teacher's target answer $y_t$ (Sec.~\ref{sec:targets}). This distinction is crucial because the soft prompt can alter the clean answer itself, for instance by answering a clean question that the base model declines. A question is counted as consistent if the model selects the same option on both inputs, or if it refuses on both. Conversely, any transition between answering and refusing, or any output that fails to produce a valid option or refusal, is marked as inconsistent. Because a model can be consistently wrong, and a system that always refuses is trivially consistent, we always report consistency alongside accuracy.

\textbf{Answer rate and over-refusal.} On clean audio, we report over-refusal, defined as the percentage of clean inputs that the model declines to answer. Under 100\% masking and $-20$~dB SNR noise, the audio evidence is either completely removed or severely corrupted to the point of being unanswerable. Under these severe refusal-target conditions, we report the answer rate, which is the percentage of questions for which the model still selects an option. For 100\% masking, any selected answer lacks audio support by design. For $-20$~dB SNR noise, heavy corruption obscures the necessary audio cues, so answering indicates a hallucination based on an unjustified guess.

\section{Results}
\label{sec:results}

\subsection{Robustness to perturbations}
\label{sec:robust}
\begin{table}[t]
\centering
\caption{Answer rates under 100\% masking and $-20$ dB SNR noise (percentage of questions answered with an option), and over-refusal on clean audio (percentage of clean questions declined), for B = Base, M = Manual, and AP = \textbf{\method}.}
\label{tab:halluc}
\resizebox{0.8\linewidth}{!}{
\setlength{\tabcolsep}{1pt}\begin{tabular*}{\columnwidth}{@{\extracolsep{\fill}}llrrrrrrrrr@{}}
\toprule
 & & \multicolumn{6}{c}{Answer rate $\downarrow$} & \multicolumn{3}{c}{Refusal $\downarrow$} \\
\cmidrule(lr){3-8}\cmidrule(l){9-11}
 & & \multicolumn{3}{c}{Silence} & \multicolumn{3}{c}{$-20$ dB} & \multicolumn{3}{c}{Clean} \\
\cmidrule(lr){3-5}\cmidrule(lr){6-8}\cmidrule(l){9-11}
Model & Bench. & B & M & \textbf{AP} & B & M & \textbf{AP} & B & M & \textbf{AP} \\
\midrule
Qwen2.5 & SAKURA & 59.1 & 41.5 & \textbf{0.5} & 32.4 & 20.1 & \textbf{3.7} & 16.2 & 24.7 & \textbf{0.0} \\
 & MMAU & 73.3 & 47.8 & \textbf{0.0} & 59.4 & 38.9 & \textbf{8.3} & 3.0 & 15.5 & \textbf{0.0} \\
 & MMAR & 35.8 & 17.4 & \textbf{0.0} & 19.5 & 9.1 & \textbf{10.2} & 14.4 & 33.4 & \textbf{0.3} \\
\midrule
Qwen3 & SAKURA & 26.0 & 26.8 & \textbf{0.1} & 11.7 & 11.6 & \textbf{3.3} & 10.3 & 14.6 & \textbf{0.0} \\
 & MMAU & 39.1 & 40.2 & \textbf{0.0} & 27.7 & 28.1 & \textbf{8.1} & 7.2 & 7.5 & \textbf{0.0} \\
 & MMAR & 17.3 & 17.1 & \textbf{0.0} & 9.3 & 8.3 & \textbf{5.7} & 13.0 & 13.5 & \textbf{0.0} \\
\midrule
AF3 & SAKURA & 100 & 100 & \textbf{0.3} & 100 & 99.9 & \textbf{9.0} & 0.0 & 0.3 & \textbf{0.1} \\
 & MMAU & 100 & 100 & \textbf{6.7} & 100 & 100 & \textbf{13.4} & 0.0 & 0.0 & \textbf{0.0} \\
 & MMAR & 100 & 99.8 & \textbf{5.3} & 100 & 100 & \textbf{5.2} & 0.0 & 0.0 & \textbf{0.0} \\
\bottomrule
\end{tabular*}
}
\vspace{-.5cm}
\end{table}

Table~\ref{tab:main} shows that the largest gains occur under adversarial injections, with more variable changes under noise and masking. Clean accuracy improves in six of nine model--benchmark pairs and decreases by only 0.1--1.2 percentage points in the remaining three. For Qwen2.5-Omni, clean-accuracy gains are accompanied by lower clean-audio refusal rates (Table~\ref{tab:halluc}). Meanwhile, AF3's clean accuracy changes by $-1.2$ to $+1.4$ percentage points relative to Base.

Under adversarial text and audio injections, \method\ increases both accuracy and clean-answer consistency across every evaluated model--benchmark pair. Under noise and masking, \method\ improves consistency in most conditions for the Qwen models while increasing accuracy in most cases, with the largest gains observed on Qwen3-Omni. In contrast, AF3 is already highly consistent under these natural perturbations without prompt tuning, and its results change marginally.

The tested \texttt{Manual} instruction fails to match \method's gains under adversarial injections. On Qwen2.5-Omni, it lowers accuracy across every condition because the model declines to answer more frequently; on Qwen3-Omni and AF3, it has a smaller impact. While both prompts share the same decoder position after audio encoding, only the soft prompt is explicitly optimized over clean and perturbed inputs.

\subsection{Answering under full masking and severe noise}
\label{sec:halluc}

\begin{table}[t]
\centering
\caption{Zero-shot transfer to OOD perturbations (test split). Each entry is the accuracy under the perturbation minus the clean accuracy of the same setting (points) / clean-answer consistency (\%), where a refusal on both inputs counts as consistent. AP = \textbf{\method}.}
\label{tab:ood}
\resizebox{0.8\linewidth}{!}{
\setlength{\tabcolsep}{0.6pt}\begin{tabular*}{\columnwidth}{@{\extracolsep{\fill}}llrrrr@{}}
\toprule
 & & \multicolumn{4}{c}{$\Delta$Accuracy vs.\ clean~/~consistency} \\
\cmidrule(l){3-6}
 & & \multicolumn{2}{c}{Choice permutation} & \multicolumn{2}{c}{Reverberation} \\
\cmidrule(lr){3-4}\cmidrule(l){5-6}
Model & Bench. & Base & \textbf{AP} & Base & \textbf{AP} \\
\midrule
Qwen2.5 & SAKURA & $+0.2$/82.0 & \textbf{$-1.2$/85.6} & $-8.5$/75.9 & \textbf{$-9.0$/79.8} \\
 & MMAU & $+1.4$/81.6 & \textbf{$-0.3$/81.9} & $-5.3$/82.8 & \textbf{$-5.9$/78.6} \\
 & MMAR & $-2.3$/69.5 & \textbf{$-1.4$/72.4} & $-7.4$/67.7 & \textbf{$-4.1$/71.7} \\
\midrule
Qwen3 & SAKURA & $-1.1$/86.0 & \textbf{$-0.6$/89.5} & $-4.3$/84.0 & \textbf{$-3.4$/86.6} \\
 & MMAU & $-0.9$/84.1 & \textbf{$-1.7$/87.0} & $-4.4$/83.4 & \textbf{$-4.5$/83.8} \\
 & MMAR & $+1.7$/75.8 & \textbf{$-0.2$/79.1} & $-6.3$/76.9 & \textbf{$-6.1$/79.6} \\
\midrule
AF3 & SAKURA & $-1.4$/74.0 & \textbf{$-0.8$/73.5} & $-11.3$/71.5 & \textbf{$-7.1$/75.6} \\
 & MMAU & $-2.5$/79.2 & \textbf{$-3.6$/79.5} & $-3.9$/85.3 & \textbf{$-3.3$/85.5} \\
 & MMAR & $+0.5$/66.6 & \textbf{$-1.4$/70.2} & $-5.8$/77.1 & \textbf{$-6.3$/75.5} \\
\bottomrule
\end{tabular*}
}
\vspace{-.3cm}
\end{table}

Table~\ref{tab:halluc} shows that \method\  reduces answer rates under both refusal-target conditions across all three LALMs. Under 100\% masking, answer rates drop to 0\%--6.7\%, while under $-20$~dB SNR noise, they fall to 3.3\%--13.4\%. Crucially, clean-audio refusal rates remain near zero ($\le 0.3\%$).

The reduction under full masking confirms that the model avoids selecting options when the audio provides no supporting evidence. Under $-20$~dB SNR noise, the lower answer rate demonstrates that the model refrains from making unjustified guesses when severe corruption makes the audio unanswerable. While the \texttt{Manual} instruction also reduces answering for Qwen2.5-Omni, it increases over-refusal on clean audio (up to 33.4\%). Because both corruption conditions are included in training, these results measure adherence to the trained refusal targets rather than zero-shot detection of missing evidence.


\subsection{Zero-shot transfer to unseen perturbations}
\label{sec:ood}
We evaluate zero-shot transfer on two unseen perturbations. The first is choice permutation, a text shift to which models are sensitive~\cite{zheng2024mcq,lopez2026robustness}. Options are reordered so that no choice keeps its original position. The second is 1-second room reverberation. Unlike noise or masking, reverberation spreads the audio signal over time without adding noise or removing content.

Table~\ref{tab:ood} shows higher clean-answer consistency in eight of nine model--benchmark pairs under choice permutation and seven of nine under reverberation. These gains are not uniform: under reverberation, consistency decreases by 4.2 percentage points for Qwen2.5-Omni on MMAU and by 1.6 points for AF3 on MMAR. Accuracy changes relative to clean performance are also mixed; for instance, \method\ reduces the reverberation-induced accuracy drop by 3.3 points for Qwen2.5-Omni on MMAR and by 4.2 points for AF3 on SAKURA. Overall, clean-answer consistency gains transfer to most OOD conditions, though the benefit depends on the model, benchmark, and perturbation.

\vspace{-.1cm}
\section{Discussion and Conclusion}
\label{sec:conclusion}
We propose \method, a parameter-efficient soft-prompting framework that adapts frozen LALMs via self-distillation and refusal targeting. Across multiple benchmarks and models, \method\ improves robustness against audio and text perturbations while preserving clean-audio performance. Furthermore, by learning to decline under severe corruption, the method successfully mitigates hallucinations without triggering false refusals on clean inputs. Finally, the learned prompt vectors demonstrate zero-shot transferability to unseen perturbations, such as choice reordering and acoustic reverberation. Overall, \method\ provides a practical and efficient approach for stabilizing multimodal language models against real-world input distortions.

\clearpage
\section{Acknowledgments}
\noindent\textbf{Funding:} Natural Sciences and Engineering Research Council of Canada (RGPIN-2023-0575: Cem Subakan; RGPIN-2022-05088: Mirco Ravanelli) and the Digital Research Alliance of Canada. Mirco Ravanelli is also supported by Translated’s Immediate research grant.

\vspace{1pt}\noindent\textbf{Ethics:} No human-subject data were collected as part of this study.

\balance
\interlinepenalty=10000
\makeatletter\let\old@thebib\thebibliography\renewcommand{\thebibliography}[1]{\old@thebib{#1}\setlength{\itemsep}{0pt}\setlength{\parsep}{0pt}}\makeatother
\bibliographystyle{IEEEbib}
\bibliography{references}
\end{document}